\documentclass{ptephy}

\usepackage{amsmath}
\usepackage{amssymb}
\usepackage{graphics}
\usepackage{epsfig}
\usepackage{graphicx}
\usepackage{color}

\newcommand \beq{\begin{eqnarray}}
\newcommand \eeq{\end{eqnarray}}
\def\simge{\mathrel{%
       \rlap{\raise 0.511ex \hbox{$>$}}{\lower 0.511ex \hbox{$\sim$}}}}
\def\simle{\mathrel{
       \rlap{\raise 0.511ex \hbox{$<$}}{\lower 0.511ex \hbox{$\sim$}}}}

\begin{document}
\title{Polarization of Direct Photons from 
Gluon Anisotropy in Ultrarelativistic Heavy Ion Collisions }

\author{
\name{Gordon Baym}{1,2},
\name{Tetsuo Hatsuda}{2,3}
\thanks{Both authors contributed equally to this work.}}

\address{
\affil{1}{Department of Physics, University of Illinois, 1110  W. Green Street, Urbana, IL 61801-3080, USA}
\affil{2}{iTHES Research Group and Nishina Center, RIKEN, Wako, Saitama 351-0198, Japan}
\affil{3}{Kavli IPMU (WPI), The University of Tokyo, Kashiwa, Chiba 277-8583, Japan }
}


\begin{abstract}

 We show how anisotropy in momentum of the gluon distribution in ultrarelativistic heavy ion collisions gives rise to polarization of direct photons produced via gluon-quark Compton scattering,  as well as by quark-antiquark annihilation into a gluon-photon pair.  We estimate the polarization asymmetry from the Compton process within a toy model where the polarized photons are produced from
  thermal gluons scattered by heavy-quark scattering centers moving with 
 Bjorken boost-invariant  flow, and find that it could be as large as 10\%.   We conclude that polarization measurements of directly produced photons can shed light on gluon pressure anisotropy in the early stages of collisions.

\end{abstract}

\subjectindex{D31,D28,B69}

\maketitle

\begin{figure}[ht]
\begin{center}
\includegraphics[scale=1.0]{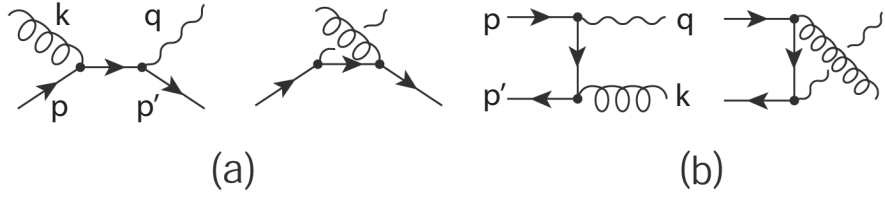} 
\end{center}
\vspace{-0.3cm}
\caption{
(a) Compton scattering of a gluon against a quark or anti-quark producing a photon; 
(b) production of a photon and gluon by quark-antiquark annihilation. 
}
\label{fig:Compton}
\end{figure}

\noindent {\em Introduction: }
 Thermal photons are a valuable probe of the state of the big-bang universe in cosmology as well as
the state of the little-bang plasma in ultrarelativistic heavy ion collisions (urHIC). 
The cosmic microwave background (CMB) provides information on the temperature of the universe at the time of decoupling 
\cite{Samtleben:2007zz}, while enhanced direct photons in urHIC \cite{Stankus:2005eq} 
provide  information on the initial temperature of the hot quark-gluon plasma 
 \cite{Adare:2008ab}.  The anisotropy of the photon spectrum ($C_l$ in cosmology and $v_n$ in urHIC) supplies further details of the initial states; 
indeed the unpolarized photon spectrum is sensitive to the 
momentum anisotropy of quarks and gluons in the evolving quark-gluon plasma 
\cite{Schenke:2006yp}.  Measurement of the CMB polarization, particularly the odd parity B-mode would, after subtraction of other sources of polarization, give critical evidence for gravitation waves created at the 
time of inflation \cite{planck}.
In this Letter we point out that the polarization of the 
direct photons from urHIC --  whose principle  sources are
Compton scattering of gluons on quarks,  Fig.~\ref{fig:Compton}(a), and quark-anti quark annihilation into a gluon-photon pair, Fig.~\ref{fig:Compton}(b),
plays a similar role, providing immediate information 
on the pre-equilibrium stage of the urHIC, especially the
deviations from local thermal equilibrium in the hot plasma.

Several distinct mechanisms of polarization of direct photons in urHIC have been considered earlier.  First, 
Ref.~\cite{Goloviznin:1997pf} predicted polarization via synchrotron
radiation of quarks in the color magnetic fields at the surface of the collision volume; however, 
whether the total yield of such photons is significant compared with the volume emission is not clear \cite{sinha}. 
Subsequently, Ref.~\cite{ipp} considered  production of circularly polarized photons
as a consequence of bulk quark polarization arising from spin-orbit coupling in non-central
collisions.   More recently, Ref.~\cite{yee}  studied photon polarization via the chiral magnetic effect,
a mechanism also effective only in non-central collisions, with the polarization asymmetry found to be 
at  the level of  0.1-0.2\%.   We stress that these varied physical mechanisms are different from 
the simple and robust  mechanisms of Compton scattering and pair annihilation we consider here.

 The local emission rate of a direct photon of four-momentum $q$ and polarization $\varepsilon_q$ from the process $1+2 \rightarrow  3 + {\rm photon}$ is generally written as  \cite{Kapusta-Gale}:
 \beq
  \frac{d{R}(q,\varepsilon_q)}{d\Gamma} 
  =\frac{{\cal D}}{2(2\pi)^3} \int \prod_{i=1}^3 \frac{d^3p_i}{2E_i (2\pi)^3}
    f_1 f_2 (1 \mp f_3) (2\pi)^4 \delta^4(p_1+p_2-p_3-q) |{\cal M}|^2, 
 \label{rgamma}
 \eeq
where $d\Gamma=E_q^{-1} d^3q$ is
 the Lorentz invariant volume element, $|{\cal M}|^2$ is the invariant matrix element squared
 with the average over 
spin, color, and gluon polarization in the initial state and the sum over final states,
 ${\cal D}$ is the degeneracy factor of the initial state, 
 and the $f$'s are the quark and gluon distribution functions. 
 For the Compton process (Fig.~\ref{fig:Compton}(a)), we take $(p_1,p_2,p_3)=(k,p,p')$ in Eq. (\ref{rgamma})
 with the minus sign in front of $f_3$, while for the pair annihilation process
 (Fig.~\ref{fig:Compton}(b)),  we take $(p_1,p_2,p_3)=(p,p',k)$ in Eq.~(\ref{rgamma})
 with the plus sign in front of $f_3$.  The degeneracy factors in Compton and annihilation  processes
  for 2-flavor QCD are
  ${\cal D}_{\rm C} = (2_{\rm spin} \times 8_{\rm color}) \times  
   (2_{\rm spin} \times 3_{\rm color} \times 2_{q\bar{q}}) \times [(2/3)^2+(-1/3)^2] = 320/3$ and 
  ${\cal D}_{\rm A} = (2_{\rm spin} \times 3_{\rm color})^2 \times [(2/3)^2+(-1/3)^2] = 20$, respectively.
      
The invariant matrix element squared with the degeneracy factor for the Compton process is \cite{ipp} 
\beq
{\cal D}_{\rm C} \overline{  |{\cal M}_{\rm C}|^2 }
 =  \frac{320}{9}e^2 g^2\left[\frac{\{qp\}}{\{qp'\}} + \frac{\{qp'\}}{\{qp\}}\right.  
  -2m^2\left.\left(\frac{\{\varepsilon_q p\}}{\{qp\}}
   - \frac{\{\varepsilon_q p'\}}{\{qp'\} }  
  \right)^2 \right];
 \label{mcbar}
\eeq
here $g$ is the QCD coupling constant, $m$ the quark mass, $e$ the electric charge,
and $\{...\}$ denotes the four-vector product. 
This formula is manifestly gauge invariant under the shift 
$\varepsilon_q \rightarrow \varepsilon_q + q$.  Also,
it reduces to the standard Klein-Nishina formula in the rest frame of the initial quark.
The final polarized photon spectrum is obtained by integration over the space-time volume of the quark-gluon plasma.

 The invariant matrix element squared for the  $q\bar q$ annihilation process can be obtained 
 from  Eq.~(\ref{mcbar})  by using crossing symmetry, $p' \rightarrow -p'$, $k \rightarrow -k$, together with an
 overall minus sign:
  \beq
{\cal D}_{\rm A} \overline{  |{\cal M}_{\rm A}|^2 }
 = \frac{160}{9}e^2 g^2 \left[\frac{\{qp\}}{\{qp'\}} + \frac{\{qp'\}}{\{qp\}}\right.  
  + 2m^2\left.\left(\frac{\{\varepsilon_q p\}}{\{qp\}}
   - \frac{\{\varepsilon_q p'\}}{\{qp'\} }  
  \right)^2 \right] .
   \label{mabar}
 \eeq

   Equations~(\ref{mcbar}) and (\ref{mabar}) are the leading order results of naive perturbation theory in which 
  $m$ is the current quark mass (a few  MeV for u and d quarks, and about 100 MeV for the s quark).
  In a strongly interacting plasma  with temperature not far from
   the critical temperature $T_{\rm c} \simeq \,$160 MeV, we would need to 
   evaluate the amplitudes by taking into account at least two effects:
   (i)  the remnant chiral symmetry breaking due to the smooth chiral crossover
     \cite{Aoki:2006we} in which the $m$'s in  Eqs.~(\ref{mcbar}) and (\ref{mabar})
     should be identified as dynamical quark masses, which are considerably
     larger than current quark masses, (ii) the effect of infrared screening cutoffs through hard thermal loops, of order $gT$. 
    We leave detailed calculations of 
   the in-medium polarization amplitudes including these effects for future studies.         
 
\begin{figure}[ht]
\begin{center}
\includegraphics[scale=0.5]{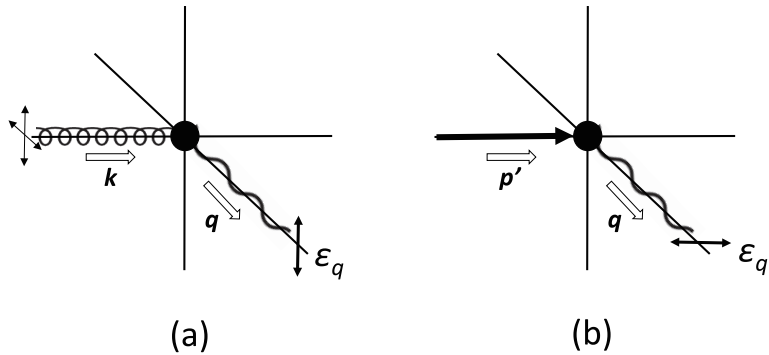} 
\end{center}
\caption{
(a) A polarized photon emitted at 90 degrees in
Compton scattering of a gluon ($k$) against a quark at rest.  
(b) A polarized photon emitted at 90 degrees in the 
pair annihilation of an anti-quark ($p'$) against a quark at rest.
}
\label{fig:Rest}
\end{figure}

 {\em Origin of the photon polarization:} 
 The sign difference in front of $2m^2$ in Eq.~(\ref{mcbar}) compared with Eq.~(\ref{mabar}) indicates that the polarizations of the photons produced in the two processes tend to be perpendicular to each other.  Explicitly, in the rest frame of the initial quark $\vec{p}=0$, in the gauge
with $\varepsilon_q$ is space-like in this frame, the polarization-dependent term
inside the bracket of Eq.~(\ref{mcbar}) reads $-2 (\vec{\varepsilon}_q \cdot \hat{k})^2$,
while that in Eq.~(\ref{mabar}) reads $+2 (\vec{\varepsilon}_q \cdot \hat{k})^2$.
The physics of difference can be seen intuitively in Fig.~\ref{fig:Rest} for photons produced at 90 degrees to the incident gluon or anti-quark.     Since the photon is produced by the
charge current in the process, its polarization is along the component of the charge current
perpendicular to the photon momentum.   A photon produced by Compton scattering of a gluon against a quark at rest is preferentially  
polarized in the direction perpendicular to the scattering plane, since the charge current is along the direction of the gluon polarization;
for a photon produced at 90 degrees, the gluon polarization out of the reaction plane is perpendicular to the photon momentum,
and thus the photon is also polarized perpendicular to the reaction plane.  
Similarly, the photon from pair annihilation of an anti-quark with a quark at rest is mostly 
polarized in the direction parallel to the scattering plane, since the 
charged current is along the incident anti-quark momentum.  (However, it is not obvious a priori which process is the dominant source of the photon polarization, since
the distribution functions $f_i$ enter differently between the two processes in Eq.~(\ref{rgamma}).)

Although the polarization-dependent and
independent terms are generally comparable in the heavy-quark limit  $m \gg q_0$, one must exercise care
in dealing with finite mass quarks.   Averages over scattering angles are most readily carried out in the 
center-of-mass frame of the incident particles, in which all final scattering angles are {\em a priori} equiprobable. 
 One would find,
for light current masses,  $m \ll q_0$, in this frame, that the scattering amplitude is dominated by backscattering, $\hat q\cdot\hat k \simeq -1$,
for which the polarization-independent terms in Eqs.~(\ref{mcbar}) and (\ref{mabar})  
are large relative to the former, suppressing the polarization asymmetry.  However owing to effects of
dynamical quark masses and infrared screening, as mentioned above, one is essentially in an intermediate mass
regime, and thus a realistic estimate of the magnitude of the polarization asymmetry must include both effects.

{\em Toy model with fixed scattering centers:}
To demonstrate as simply as possible the essence of the photon polarization produced in heavy ion collisions
 we    focus on the Compton process and schematically regard the quarks as heavy scattering
 centers co-moving with the fluid.   This approximation, although drastic for the
 quarks in the thermal medium, highlights the role of the 
 initial anisotropy of the gluon distributions in producing polarized photons. 
 In this model, we take the quark momentum distribution in the rest frame of the fluid to be proportional to
  $\delta^3(\vec{p})$,
and take the quark mass $m$ to infinity in $|{\cal M}|^2$.
 Then the emission rate ({\ref{rgamma}) becomes 
 \beq
  \frac{d{R}(q,\varepsilon_q)}{d\Gamma} 
  = {\cal N} n(x)    \int \frac{d\Omega_k}{4\pi} f_{{\vec k}}(x) \left( 1- (\vec \varepsilon_q \cdot \hat k)^2 \right),
  \label{emm}
\eeq  
where all momentum independent factors are included in the constant ${\cal N}$,
 $n(x)$ is the quark (and anti-quark) number density in the local fluid element at $x$, and $f_{\vec k}$ 
 with $k=|\vec k|=|\vec q|=q$ is the (time-dependent)
 gluon momentum distribution in momentum space.
 If the gluon distribution is isotropic in $\vec k$, then simply 
 $\langle (\vec{\varepsilon}_q \cdot \hat k)^2 \rangle =  \frac{1}{3}$, where 
 \beq
\langle {\cal O} \rangle \equiv  \frac{1}{f_k^{\rm av}}
  \int \frac{d\Omega_{\hat k}}{4\pi} f_{\vec k} {\cal O},\quad f_k^{\rm av}(x) = \int \frac{d\Omega_{\hat k}}{4\pi}  f_{\vec k},
 \eeq
denotes the angular average over $\hat{k}$ with weight factor $f_{\vec k}$.

\begin{figure}[h]
\begin{center}
\includegraphics[scale=0.55]{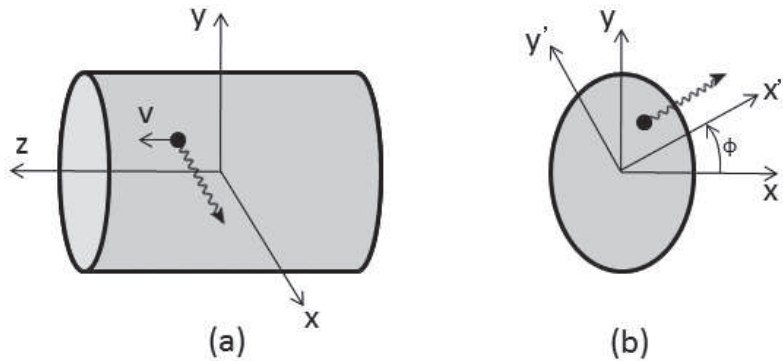} 
\end{center}
\vspace{-0.6cm}
\caption{Assumed collision geometry with beam axis in the z-direction; (a) in a central collision we consider a photon emitted in the
x-direction at central rapidity, and (b) more generally at azimuthal angle $\phi$ in the transverse plane.}
\label{fig:geometry}
\end{figure}

{\em Polarization asymmetry in central collisions:}
We first consider a central collision along the z-axis (Fig.~\ref{fig:geometry}a).
A photon emitted with momentum $\vec q$  in the rest frame of a local fluid element will appear in the laboratory frame
 to have momentum
 $ q^L= \gamma (q+ v q_z), \ \  q^L_z = \gamma(q_z + v q), \ \  \vec q\,^L_\perp = \vec q_\perp$,
where $v$ is the velocity of the fluid element in the 
laboratory frame and  $\gamma = 1/\sqrt{1-v^2}$. 
The photon yield in the laboratory is an integral of the 
emission rate over the space-time evolution of the plasma:
\beq
\frac{dN}{d\Gamma}= \int d^4x \frac{dR}{d\Gamma},
\eeq
where $d\Gamma = q_{\perp} dq_{\perp} dY d\phi_q$  with $q_{\perp}$ the  
transverse momentum and $Y$ the momentum rapidity in the laboratory frame.  In terms of the proper time
$\tau = \sqrt{t^2-z^2}$ and the space-time rapidity $\eta = \frac12\ln\left[ (t+z)/(t-z)\right]$,
  one has $d^4x = d^2 r_{\perp} \tau d\tau d\eta$.
 We consider first a photon emitted at central rapidity ($Y=0$ or equivalently $q_z^L=0$)
  in the $x$-direction.  
The polarization of the observed photon is then in the $(y,z)$ plane.
We define the {\em polarization anisotropy} \cite{yee-def}, 
\beq
 r(q_{\perp}, Y=0)
 \equiv 
 \frac{dN_{\hat{z}}/d\Gamma-dN_{\hat{y}}/d\Gamma}{dN_{\hat{z}}/d\Gamma+dN_{\hat{y}}/d\Gamma } ,
\eeq 
 where $N_{\hat{z}}$ and $N_{\hat{y}}$ are the photon yields with plane polarization along the
  $z$ and $y$ directions. 
Non-vanishing $r$ for direct photons 
signals anisotropy of the momentum distribution in the plasma.

  The photon polarization depends on the quadrupolar anisotropy of the gluon distribution, as we now show.
 A photon in the $x$-direction at $Y=0$ in the lab frame
must be emitted at an angle $\theta$ with respect to the $z$-axis in the local fluid rest frame, where 
$\cos\theta = -v$.
  If the  photon is polarized along  $\hat y$, then $\vec{\varepsilon}_q\cdot \hat k = \hat k_y$.
On the other hand, if the photon in the lab is
polarized along $\hat z$, then in the local rest  frame the photon is polarized in the $x$-$z$ plane, and 
$\vec \varepsilon_q\cdot \hat k = \sin\theta\, \hat k_z - \cos\theta \,\hat k_x$. 
 When $(\vec \varepsilon_q\cdot \hat k)^2$ is averaged over angles, the cross term  $\sim \hat k_z \hat k_x$ vanishes  by axial symmetry in the collision, and
 \beq
\frac{dR_{\hat{z}}}{d\Gamma} 
&=& {\cal A}
 \left( 1 - \sin^2 \theta \langle \hat{k}_z^2 \rangle - \cos^2 \theta \langle \hat{k}_x^2 \rangle \right) 
 = \frac{\cal A}{3} \left(2 + (1-\frac{3}{2} v^2) \lambda \right),  \label{Rz} \\
\frac{dR_{\hat{y}}}{d\Gamma} 
&=& {\cal A} \left( 1 - \langle \hat{k}_y^2 \rangle  \right)=\frac{\cal A}{3} \left(2 - \frac\lambda2 \right),
 \label{Ry}
 \eeq
where ${\cal A} = {\cal N}'n(x) f_k^{\rm av}(x)$.  The key microscopic parameter characterizing the {\em momentum anisotropy\,} is
\beq
   \lambda \equiv \langle \hat k_x^2 + \hat k_y^2 \rangle -2 \langle \hat k_z^2 \rangle ,
\eeq  
where $-2\le \lambda \le 1$.
 
The polarization anisotropy to leading order in $\lambda$ is thus
\beq 
r(q_{\perp},Y=0) 
=\frac{3}{8}\  \overline{\lambda (1-v^2)},
\label{eq:pol-aniso}
\eeq
where $\overline{\cal O}$ denotes the space-time average,
\beq
  \overline{\cal O} = \frac{ \int \tau d\tau  \int d\eta \ n(\tau) f_k^{\rm av}(\tau, \eta){\cal O} }
 { \int \tau d\tau \int d\eta \ n(\tau) f_k^{\rm av}(\tau, \eta)}.
\eeq
Note that  $1-v^2 = 1/\cosh^{2} \eta$, and that $f_k^{\rm av}(\tau, \eta)$ can be taken as isotropic in the leading order estimate of $r$.


    As a specific model of anisotropy, we consider Romatschke and Strickland's  phenomenological momentum anisotropy distribution in the local rest frame
\cite{Romatschke:2003ms},
\beq
f_{\vec k;\xi} = e^{ - [k_{x}^2+k_{y}^2 + (1+\xi) k_z^2]^{1/2}/T}.  
\label{eq:RSF}
\eeq
The parameter $\xi$, which controls the effective temperature of gluons along $z$-direction,
 $T_z =T/\sqrt{1+\xi}$, is related to 
our model-independent momentum anisotropy parameter by $\lambda \simeq 2k\xi/15T$ 
to leading order in $\xi$.  In general, $\xi$, as well as $T$, depends on space and time.

    We average over local fluid velocities, at fixed  photon lab energy $q_{\perp}$ at mid-rapidity,
within the Bjorken model of Lorentz invariant evolution \cite{bj}, in which  the local fluid frame velocity is $v = \tanh \eta$,  and the particle density $n(\tau)$ scales as $1/\tau$.
The collision volume extends in space-time rapidity $\eta$ from $-\eta_0$ to $\eta_0$.
To calculate space-time averages, one integrates over the space-time volume
 from the initial proper-time $\tau_0$ (corresponding to the initial temperature $T_0$)
to the final proper time $\tau_{\rm f}$ (corresponding to the QCD critical temperature $T_{\rm c}\sim 160$ MeV).

  Since $k = q = q_\perp\cosh\eta$ at $q_z^L=0$, , the momentum anisotropy for given $ q_\perp$ is
 \beq    
\lambda = \frac{2q_{\perp}}{15T(\tau)} \xi(\tau) \cosh \eta,
\label{xilambda}
\eeq
and is dependent on $\eta$, $\tau$ and  $q_{\perp}$. 
For a direct photon with $q_{\perp} \sim 1-3$ GeV relevant to urHIC \cite{Adare:2008ab}, 
we can safely use the Boltzmann distribution for the gluons of energy $k$ 
 in the local fluid rest frame, $
  f_k^{\rm av}(\tau, \eta) \propto \exp(-q_\perp\cosh\eta/T)$,
rather than a full Bose distribution.  Since quite generally the gluon energy required to
produce a photon of energy $q_\perp$ in the lab increases with $\eta$, the falloff of $f_k^{\rm av}$ with $k$ limits the range
of $\eta$ producing the photon.  In integrating over space and time, we can (except when considering photons produced at forward or backward rapidities) thus extend the limits of the  $\eta$ integral to $\pm\infty$.  Because the range of $\eta$ responsible for polarization of a
photon of given rapidity is limited, the rapidity distribution has a plateau about central rapidities; calculating the polarization at forward, or backward, rapidities requires keeping the upper, or lower, limit of $\eta$ finite.

  With the estimate (\ref{xilambda}) in Eq.~(\ref{eq:pol-aniso})
and the assumption that $\xi$ and $T$ depend only on $\tau$, we have 
for Compton scattering,
\beq
r= \frac{1}{20} 
\frac{\int d\tau\, \xi Q\, {\cal I}_{-1}(Q) }
{\int d\tau\,  {\cal I}_0(Q)},
\label{r}
\eeq 
where  $Q\equiv q_{\perp}/T(\tau)$ and  
\beq
  {\cal I}_n(Q) \equiv \frac{1}{2} \int_{-\infty}^{+\infty} d\eta \, e^{-Q \cosh \eta }(\cosh\eta)^n.
\eeq
In terms of $K_n(x)$ (the modified Bessel function of the second kind) and 
$L_n(x)$ (the modified Struve function), we have  ${\cal I}_{0}(x) = K_0(x)$,   ${\cal I}_{1}(x) = K_1(x)$, and
$
{\cal I}_{-1}(x) = (\pi/2) - \int_0^x dy\, K_0(y) 
= (\pi/2) \left(1- x\left(K_0(x)L_{-1}(x) + K_1(x) L_0(x)\right)\right).$

   For Bjorken flow with $T=T_{0} (\tau_0/\tau)^{1/3}$, the $\tau$ integrals
in Eq.~(\ref{r}) can be easily done by changing variables from $\tau$ to $T$.
Figure \ref{fig:rQ} shows the resulting polarization asymmetry $r$ as a function
of $q_{\perp}$ for three characteristic behaviors,  $\xi=\xi_0 \left( \tau_0/\tau \right)^{\alpha}$,
 with $\alpha$ = 0, 1/3, and 1.  We carry out the $T$ integration
for $0.3\, T_0< T < T_0 $;  the result is, however, insensitive to the choice of lower end of the integral.

\begin{figure}[h]
\begin{center}
\includegraphics[scale=0.7]{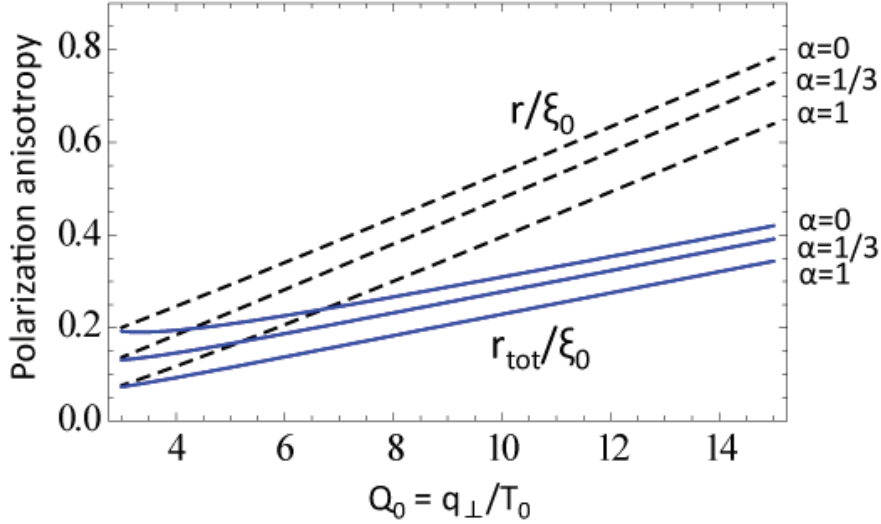} 
\end{center}
\vspace{-0.6cm}
\caption{(Color online) The Compton polarization anisotropy $r$ (dashed lines) 
and the total polarization anisotropy $r_{\rm tot}$ (solid lines) 
 for a photon at central rapidity of momentum $q_\perp$, normalized by the initial anisotropy $\xi_0$,
 as a function of  $Q=  q_\perp/T_0$ for $\alpha$ = 0, 1/3 and 1;  $T_0$ is the initial temperature. }
\label{fig:rQ}
\end{figure} 
 
As Fig.~\ref{fig:rQ} shows, $r$ (dashed lines) increases  linearly with $q_{\perp}$. 
The source of the linearity is that
 ${\cal I}_{-1}(Q)$ and ${\cal I}_0(Q)$ in Eq.~(\ref{r}) have the same asymptotic behavior: ${\cal I}_{-1,0,1}(x) \rightarrow \sqrt{\pi/2x} \,\,e^{-x}$;
thus in the limit $\tau_{\rm f}\rightarrow \tau_0$, we obtain the analytic relation, $r= \xi Q/20$.  The integration over $\tau$ does not change this 
 feature significantly. The polarization anisotropy measured with respect to the total direct photon yield can be written approximately as
\beq
r_{\rm tot}(q_{\perp},Y=0) \sim 0.5\,  f_{\rm C}\, \xi_0\, 
\left( \frac{q_{\perp}}{2\ {\rm GeV}} \right) \left( \frac{0.2\ {\rm GeV}}{T_0} \right),
\label{eq:r_central}
\eeq
where $f_{\rm C}  \simeq \rho_{_{\rm C/P}}/(1+\rho_{_{\rm C/P}})$ is the fraction of the direct photons produced by Compton scattering,
and $\rho_{_{\rm C/P}}$ is 
 the ratio of the rates of unpolarized photon emission from Compton to pair annihilation;  for massless quarks
 \cite{Kapusta-Gale}, 
 $ \rho_{_{\rm C/P}} = 0.5 (\ln (12Q_0/g^2)+0.046)(\ln (12Q_0/g^2)-2.147)$.
 With $g^2 = 4$, this ratio decreases with $Q_0$ from $\simeq$ 2.5 for $Q_0$ = 5, where $f_{\rm C} = 0.71$, to $\simeq$ 1.4 
 for $Q_0$ = 10, where $f_{\rm C} = 0.58$. 
As Fig.~\ref{fig:rQ} shows,  $r_{\rm tot}$ (solid lines) decreases more slowly with $Q_0$ due to the relative decrease of Compton photons.
 Moderate values of the averaged  anisotropy, $\xi_0 \sim 1$, give $r_{\rm tot}
\sim$ 10-40 \%.  As we saw earlier, calculating with finite quark masses will reduce this estimate by a factor $\sim 2$.

 {\em Polarization asymmetry in non-central collisions:}
 We next consider photon polarization at central rapidity in non-central collisions.  We work in the usual frame in which the reaction plane is the $x$-$z$ plane.
 A photon emerging at angle $\phi$ with respect to the $x$ axis can be polarized
 in the $z$-$y'$ plane  (Fig.~\ref{fig:geometry}b) where the $y'$-axis is at  angle $\phi+ \pi/2$ to the x-axis.  Allowing for 
 momentum anisotropy in the transverse  x-y plane 
 in a non-central collision, we introduce the transverse momentum anisotropy parameter,
 \beq
   \nu =  \langle \hat k_x^2\rangle - \langle \hat k_y^2\rangle .
\eeq
 We assume that $\langle \hat k_i \hat k_j\rangle$ is diagonal in the present coordinate system.
 Then $ \langle \hat k_x^2\rangle =\frac{1}{3}( 1+\lambda/2) + \nu/2$, 
$\langle \hat k_y^2\rangle = \frac{1}{3}( 1+\lambda/2) - \nu/2$, and
  $ \langle \hat k_z^2\rangle =\frac{1}{3} (1-\lambda)$.

 If the  photon is polarized along the $y'$-axis, then $\vec{\varepsilon}_q\cdot \hat k = 
 \cos \phi\, \hat k_y  - \sin \phi\,  \hat k_x $, and $\langle (\vec\varepsilon\cdot \hat k)^2\rangle = 
 \frac13(1+\lambda/2) -(\nu/2)\cos2\phi$, so that the emission rate is [cf. (\ref{Ry})], 
 \beq
 \frac{dR_{\hat{y'}}}{d\Gamma} =\frac{\cal A}{3} \left(2 - \frac\lambda2  + \frac{3}{2}  \nu \cos 2\phi\right). 
 \eeq
On the other hand, if the photon in the lab is polarized along the z-axis,  
$\vec \varepsilon_q\cdot \hat k = \sin\theta\, \hat k_z - \cos\theta \,\hat k_{x'}$
 with $\hat k_{x'}=  \cos \phi\, \hat k_x  + \sin \phi\,  \hat k_y $, and $\langle( \vec\varepsilon\cdot\hat k)^2\rangle =
 \frac13\left(1+\lambda(\frac32\cos^2\theta - 1)\right) + \frac12\nu\cos^2\theta\cos2\phi$; thus
the emission rate is
\beq
   \frac{dR_{\hat{z}}}{d\Gamma} 
=  \frac{\cal A}{3}\left(2 + \left(1-\frac32{v^2}\right)\lambda - \frac{3}{2} {v^2} \nu\cos 2\phi  \right).
\eeq   

To first order in $\lambda$ and $\nu$ the polarization anisotropy of Compton photons [cf. (\ref{eq:pol-aniso})] 
generalizes to
\beq 
r(q_{\perp},Y=0) 
=\frac{3}{8}\left( \overline{\lambda (1-v^2)} -  \overline{\nu(1+v^2)}\, \cos2\phi\right). 
\label{eq:pol-anisonu}
\eeq

  To estimate the polarization produced by the azimuthal anisotropy,
we generalize Eq.~(\ref{eq:RSF}) to $ f_{\vec k;\zeta,\xi}= \exp\left(- [(1-\zeta )k_{x}^2+(1+\zeta)k_{y}^2 + (1+\xi) k_z^2]^{1/2}/T\right)$,  
in terms of which $\lambda \simeq 2k\xi/15T$ and $\nu \simeq 2k\zeta/15T$ to leading order.
Then averaging over space and time we find
\beq
r=\frac{\int d\tau Q \left[\xi{\cal I}_{-1}(Q)  - \zeta\cos2\phi 
 (2{\cal I}_{1}(Q)-{\cal I}_{-1}(Q))\right]}{20 \int d\tau\,{\cal I}_0(Q)} . 
\eeq 
For $T$ scaling as $\tau^{-1/3}$ together with the asymptotic form of ${\cal I}_{-1,0,1}(Q)$,
we obtain the same approximate formula as Eq.~(\ref{eq:r_central}) 
with the replacement, $\xi_0 \rightarrow \xi_0 - \zeta_0 \cos 2\phi$.
If  $\zeta$ is the same order as $\xi$,  the longitudinal and axial anisotropies lead to comparable photon polarization, and  their relative contributions can be distinguished by the $\phi$ dependence of the polarization.

 The anisotropy of the gluon distribution is reflected as well in the anisotropy 
 of the local gluon stress tensor, 
$T_{ij} = 16\int_{\vec k} {k_ik_j}f_{\vec k}$.
In terms of $\lambda$ and $\nu$,
   $T_{xx}+T_{yy}-2T_{zz} = 16\int_k k^2 f_k^{\rm av} \lambda(k)$, and  
   $T_{yy}-T_{xx} =16\int _k k^2 f_k^{\rm av} \nu(k)$.  
The processes that give rise to anisotropy of the hydrodynamic evolution also give rise to photon polarization.
Although the former depends on the anisotropy of the  gluon distribution at lower momenta than does the direct photon polarization, the same parameters $\lambda(k)$ and $\nu(k)$ enter.

 {\em Summary:} 
    As shown in this Letter,
measurement of the polarization of direct photons in urHIC is a valuable probe of
the anisotropy of the gluon distribution in the evolving quark-gluon plasma.  
 The present estimates of the relation between the polarization anisotropy, $r$, 
   and the momentum anisotropies, $\lambda$ and $\nu$, 
 suggest the feasibility of detecting the photon polarization in heavy-ion collisions,  e.g.,
via the conversion of thermal photons to $e^+e^-$ 
with subsequent measurement of the angular  distribution of the lepton pairs \cite{exp-ack}.  Optimal would be to look at lower transverse momenta, $p_{\rm T} \sim$ 1 GeV,  where 
thermal photons dominate over the perturbative QCD background
(see Fig.~4 of Ref.~\cite{Adare:2008ab}), and simultaneously the opening angle of the converted lepton pair is not too small.   A full analysis with an appropriate experimental setup is, however, beyond the scope of this paper.

 In a future publication we will provide detailed calculations of the expected polarization distributions, with full kinematics of Compton scattering and pair annihilation, including detailed
quark and antiquark distributions in the evolution of the collision \cite{BIHS}. 
 Further topics to be addressed are a more complete treatment of the gluons including effects of initial gluon polarization, longitudinal gluons, and finite screening masses; polarization effects of virtual photons; and deviations from the plateau in polarization at forward and backward rapidities.  A fully quantitative calculation of the polarization must also take into account the effects in Refs. \cite{Goloviznin:1997pf}, \cite{ipp}, and \cite{yee}.   A better understanding of the origin of momentum anisotropy in terms of the kinetic evolution of non-equilibrium distributions from the initial state is also needed for a complete picture of polarization of direct photons in ultrarelativistic heavy ion collisions.

\section*{Acknowledgments}
   We both thank the RIKEN iTHES project for partial support during the course of this research.   This research was also supported in part by NSF Grants PHY09-69790 and PHY13-05891, and JSPS Grants-in-Aid No.~25287066.

\vspace{100pt}


\begin{thebibliography}{9}
\bibitem{yee-def}  This definition is essentially that introduced in Ref.~\cite{yee}.
\end{thebibliography}
\end{document}